\documentclass[a4paper,twocolumn,superscriptaddress,11pt,accepted=2017-08-14]{quantumarticle}
\pdfoutput=1
\usepackage[utf8]{inputenc}
\usepackage[english]{babel}
\usepackage[T1]{fontenc}
\usepackage{amsmath}
\usepackage{hyperref}

\usepackage{amsmath,amsthm,amsfonts,amssymb,times,graphicx,color}
\usepackage{multirow}
\usepackage{subfigure}
\usepackage{mathtools}
\usepackage{url}
\usepackage{color}
\mathtoolsset{showonlyrefs}

\renewcommand{\Re}{\operatorname{Re}}
\renewcommand{\Im}{\operatorname{Im}}

\usepackage{amsmath,amsfonts,amssymb,bbm}

\newcommand{\bra}[1]{\langle #1 |}
\newcommand{\braket}[2]{\langle #1 | #2 \rangle}

\newcommand{\ket}[1]{| #1 \rangle}

\newcommand{\couic}[1]{}

\newcommand{\C}{\mathbb{C}}

\newcommand{\Z}{\mathbb{Z}}

\newcommand{\ii}{\mathrm i}

\def\PA#1{{\textcolor{magenta}{}}}

\usepackage{tikz}
\usepackage{lipsum}

\begin{document}

\title{Quantum walking in curved spacetime: discrete metric}
\date{\today}
\author{Pablo Arrighi}
\email{pablo.arrighi@univ-amu.fr}
\affiliation{Aix-Marseille Univ, Universit{\'e} de Toulon, CNRS, LIS, Marseille, and IXXI, Lyon, France}
\author{Giuseppe Di Molfetta}
\email{giuseppe.dimolfetta@lis-lab.fr}
\homepage{http://www.giuseppe-dimolfetta.com}
\affiliation{Aix-Marseille Univ, Universit{\'e} de Toulon, CNRS, LIS, Marseille, France  and Departamento de F{\'i}sica Te{\'o}rica and IFIC, Universidad de Valencia-CSIC, Dr. Moliner 50, 46100-Burjassot, Spain}
\author{Stefano Facchini}
\email{stefano.facchini@univ-amu.fr}
\affiliation{Aix-Marseille Univ, Universit{\'e} de Toulon, CNRS, LIS, Marseille, France}

\maketitle

\begin{abstract}
A discrete-time quantum walk (QW) is essentially a unitary operator driving the evolution of a single particle on the lattice. Some QWs have familiar physics PDEs as their continuum limit. Some slight generalization of them (allowing for prior encoding and larger neighbourhoods) even have the curved spacetime Dirac equation, as their continuum limit. In the $(1+1)-$dimensional massless case, this equation decouples as scalar transport equations with tunable speeds. We characterise and construct all those QWs that lead to scalar transport with tunable speeds. The local coin operator dictates that speed; we provide concrete techniques to tune the speed of propagation, by making use only of a finite number of coin operators---differently from previous models, in which the speed of propagation depends upon a continuous parameter of the quantum coin. The interest of such a discretization is twofold : to allow for easier experimental implementations on the one hand, and to evaluate ways of quantizing the metric field, on the other. 
\end{abstract}

\section{Introduction}

{\em Discrete space discrete time quantum mechanics\ldots} Physics traditionally describes physical phenomena in terms of Partial Differential Equations (PDE). Next in order to convert this description, into a way of making predictions, the PDE gets discretized in space and time, through the application of different numerical methods. Typically, these discretization methods break the symmetries and the conservation laws of the continuous theory. The unitarity of quantum mechanics gets broken, for instance, whenever a quantum physical phenomenon gets discretized through Finite-Differences. Indeed, although unitarity provides numerical stability \cite{ArrighiDirac}, if the aim is just to run the simulation on a classical computer, there may be easier ways. 

{\em \ldots for quantum simulation\ldots} But quantum mechanics is notoriously expensive to simulate on a classical computer. Just the size of the classical description of the wavefunction grows exponentially in the number of quantum systems that one wishes to simulate. Faced with this issue, Feynman realized \cite{FeynmanQC} that in order to simulate a quantum mechanical system we ought to use\ldots another quantum mechanical system. In his view, a quantum computer is therefore just a well-controlled, fine-tunable quantum device, which one can use to mimic different quantum physical phenomena. Other applications of Quantum Computing have been invented since, but these require so far unreachable implementation precisions \cite{Steane}. Quantum simulation devices, on the other hand, are being implemented  \cite{WernerElectricQW,Sciarrino} and are likely to be useful even with some noise. Now consider a given quantum physical phenomenon, as described by a PDE. Discretizing it through the traditional numerical methods will typically fail to produce a quantum simulation algorithm runnable on quantum simulation device. Unitarity, for instance, is now a necessity---as the device itself obeys quantum mechanics. But there are other criteria for a good quantum simulation algorithm: that it preserves the space-time structure of the phenomenon (for parallelism, and so that its noise model maps onto the device's noise); that it uses a minimal number of gates (for ease of implementation, or efficiency).  

{\em \ldots or as theoretical physics toy models.} Finding a good quantum simulation algorithm, has much to do with finding a fully legitimate discrete space, discrete time quantum mechanical description of the phenomenon. Such a description may have an interest per se, whether because the phenomenon is in fact suspected to be discrete at a smaller scale (e.g. Planck scale) or because phrasing it in discrete terms makes its simpler, more explanatory.\\
For these two reasons, the recent years have witnessed an explosion in the number of discrete space-time quantum algorithmic descriptions of physical phenomena---mainly through Quantum Walks. Of particular interest to us are those which lie at the intersection of relativity and quantum mechanics, both for theoretical reasons (this part of theoretical physics remains to be fully understood) and practical reasons (this phenomena are hard to experiment with).

{\em Quantum Walks in Curved Spacetime.} A discrete-time Quantum Walk (QW) is essentially an operator driving the evolution of a single particle on the lattice, through local unitaries. Whilst some Quantum Computing algorithms are formulated in terms of QWs, see \cite{venegas2012quantum}, we focus here on their ability to simulate certain quantum physical phenomena, in the continuum limit. After it became clear that QWs can simulate the Dirac equation \cite{BenziSucci, Bialynicki-Birula, MeyerQLGI,LapitskiDellarPalpacelliSucci,DAriano,ArrighiDirac}, the Klein-Gordon equation \cite{IndiansDirac,ArrighiKG, MolfettaDebbasch} and the Schr\"odinger equation \cite{StrauchShrodinger,LoveBoghosian}, the focus moved towards simulating particles in some background field \cite{cedzich2013propagation, di2014quantum, marquez2017fermion, di2016quantum, arnault2016quantum}, with the difficult topic of interactions initiated in \cite{meyer1997quantum, ahlbrecht2012molecular}. The question of the impact, of these inhomogeneous fields, upon the propagation of the walker gave rise to lattice models of Anderson localization \cite{WernerLocalization,JoyeLocalization}. Surprisingly, it also gave rise to lattice models of particles propagating in curved spacetime. The original work on simulating the Weyl equation in $(1+1)-$dimensions in synchronous coordinates \cite{di2013quantum,di2014quantum}, was later extended \cite{ArrighiGRDirac} to the following entire class of equations
\begin{small}
\begin{equation}
\partial_t \psi(t,x) = B \partial_x \psi(t,x) + \frac{1}{2} \partial_x B \psi(t,x) + \ii C\psi(t,x), \label{eq:ContLimitQW}
\end{equation}
\end{small}
where $B$ and $C$ are two (possibly space-time dependent) hermitian matrices and $|B|\leq I$. In particular, notice how the eigenvalues of $B$ are able to tune the speed of the underlying equations anywhere between $-1$ and $1$ according to whatever the metric demands. Notice also the second term, which is necessary for the probability conservation, and the last, which can code both for a mass term or an effective field interaction. This class indeed includes the hamiltonian form of the Dirac equation in $(1+1)-$dimensional curved spacetime in arbitrary coordinates and in the presence of an electric field \cite{de1962representations}. This was eventually generalized to arbitrary spatial and spin dimensions in \cite{DebbaschWaves,ArrighiGRDirac3D}. Let us notice that some of the authors have rigorously proved that the solutions of the QW, even in $(3+1)-$ dimensions, observationally converge toward the solutions of the Dirac Equation provided the initial conditions are smooth enough,
as quantified by a Sobolev norm \cite{ArrighiDirac}.

{\em Discretizing the coin, quantizing the metric.} All of these models \cite{cedzich2013propagation,di2014quantum,ahlbrecht2012molecular,WernerLocalization,
JoyeLocalization,di2013quantum,ArrighiGRDirac,DebbaschWaves, marquez2017fermion,ArrighiGRDirac3D,di2016quantum} are inhomogenous QW. They work by computing the lattice wavefunction at $\psi(t+\varepsilon)$ from that at $\psi(t)$ by applying local unitary matrices $W(x,t)\in U(2k)$ across space (see Fig. \ref{fig:LLLat2}). These local unitary matrices or `coins', are potentially space-time dependent; in the context of \cite{di2013quantum,di2014quantum,ArrighiGRDirac,
DebbaschWaves,ArrighiGRDirac3D} they depend on the metric $g(x,t)$. In fact, they depend continuously upon $g$, which implies that the set $\{W(g)\}_g$ is in principle infinite. This makes it difficult to implement. Experimentally, one would prefer the set of possible coins to be finite. Hence, the question: {\em Is it possible to simulate Eq. \eqref{eq:ContLimitQW} with a finite number of coins?}\\
This question, it turns out, echoes with deeper concerns in theoretical physics. Indeed, most modern approaches towards building a Quantum Gravity theory argue that \cite{RovelliLQG,Bekenstein,nicoletopoulous1990two} the geometry of spacetime should be ultimately discrete in order to be quantized---and take this as their point of departure \cite{LollCDT,QuantumGraphity1}. The ability to simulate Eq. \eqref{eq:ContLimitQW}, for an arbitrary $B$, with just a finite number of coins, would make a strong mathematical argument in favor of an underlying discrete structure from which propagation in continuous curved spacetime can emerge. We evaluate this question. In order to give a general, but tractable treatment, we focus on the massless case of \eqref{eq:ContLimitEq} -- i.e. the Weyl Equation (WE), looking at the dynamics of one of the spinor components as it propagets in $(1+1)-$dimensional curved spacetime: 
\begin{equation}
\partial_t\psi =c\partial_x\psi+\frac{1}{2}(\partial_x c)\psi.
\label{eq:scalarpropagation0}
\end{equation}
We seek to obtain this equation as the continuum limit, when $\varepsilon\rightarrow 0$, of an $\varepsilon-$parameterized discrete space discrete time evolution \`a la:
\begin{equation}
\ket{\psi(t+4\varepsilon)}=(\bigoplus_{\varepsilon 2\mathbb{Z}+\varepsilon} W_\varepsilon)(\bigoplus_{\varepsilon 2\mathbb{Z}} W_\varepsilon)\ket{\psi(t)},\label{eq:evolint}
\end{equation}
with $W$ a spacetime dependent $U(k)$, and up to a unitary encoding and grouping of the field $\psi$. We provide a number of positive and negative results.   

{\em Plan, results.} Section \ref{sec:model} describes the discrete model, i.e. the rather wide class of QWs that we will consider. It sets up the notation. Section \ref{sec:fixedspeedconditions} takes the continuum limit of these wide class of quantum walks, in their full generality, and derives necessary conditions so that they may have \eqref{eq:ContLimitQW} as their continuum limit---with arbitrary yet fixed speed of propagation at this stage. Section \ref{sec:fixedspeedconditions} shows how to construct all possible solutions of these constraints. It also shows that several speeds can be implemented by a single $W$. However, it shows that these speeds do not form a continuum---a single $W$ can only implement a finite number of speeds. Section \ref{sec:discretemetric} shows how to circumvent this problem: two simple $W_0$ and $W_1$ can be combined, to make up for an infinity of possible larger $W$s, each capable of handling a certain speed. This `discretized metric' construction is very natural: the simulation remains real-time, and background signals travelling at lightspeed may be used to control which of $W_0$ or $W_1$ should be applied. Section \ref{sec:nonfixedspeedcaseconditions} generalizes Section \ref{sec:fixedspeedconditions} to look at the conditions for implementing \eqref{eq:ContLimitQW} with varying speed of propagation. It shows that this varying speed of propagation needs be differentiable if we are to recover the traditional curved spacetime transport equation. Section \ref{sec:nonfixedspeeddiscretizedmetric} shows that the discretized metric scheme does not meet this differentiability condition. In some interesting lightlike-splitted cases this can be fixed by means of local rotations that seem suitable for quantum simulation. Section \ref{sec:conclusion} summarizes the results. It discusses the generality of the constructions and obstructions found, and opens new perspectives. 

\section{The Model}\label{sec:model}

\begin{figure}
\includegraphics[width=\columnwidth]{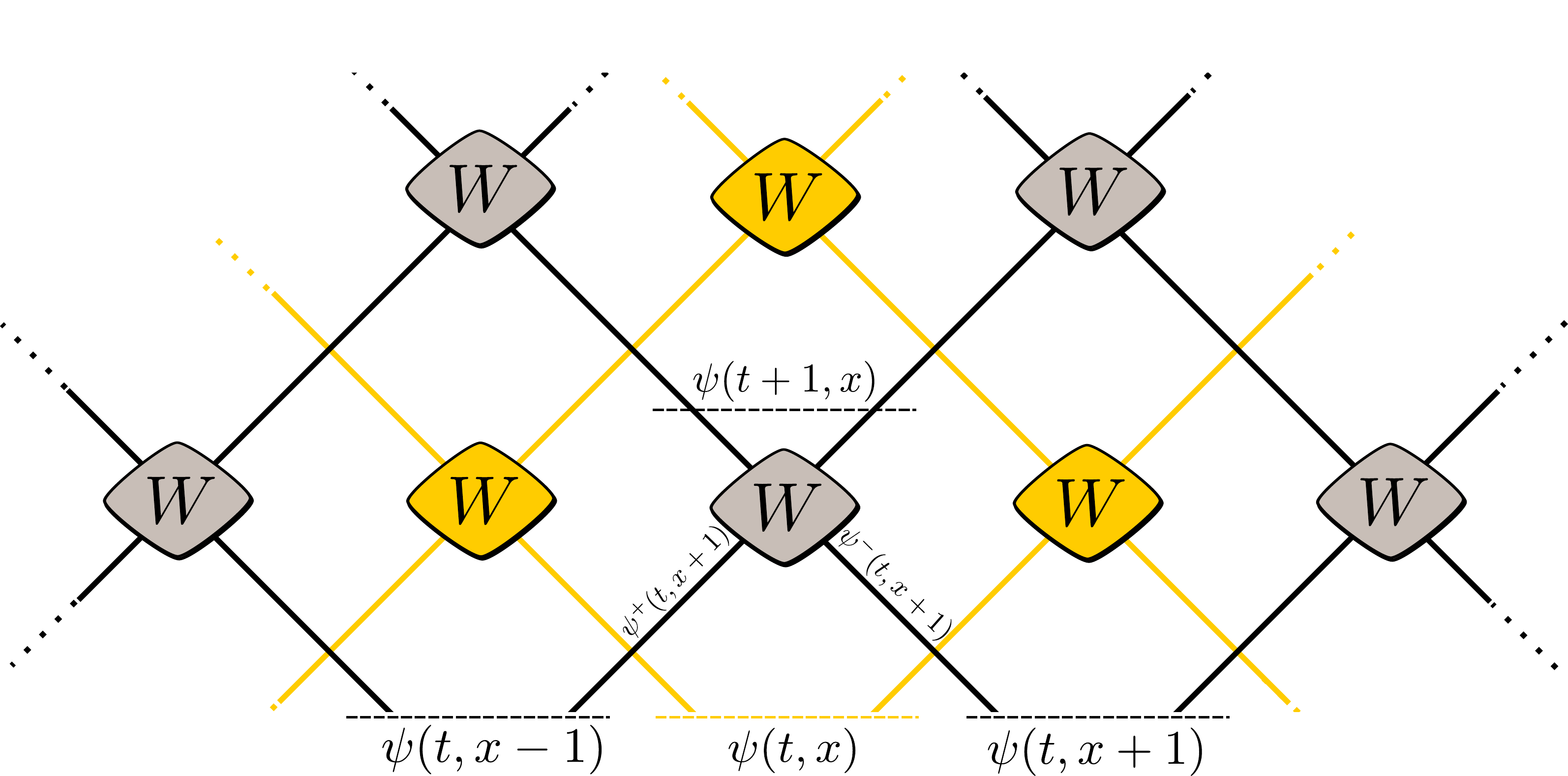}
    \caption{(Color Online) Usual QWs. Times goes upwards. Each site contains a $2d$-dimensional vector $\psi=\psi^+\oplus\psi^-$. Each wire propagates the $d$-dimensional vector $\psi^\pm$. These interact via the $2d\times 2d$ unitary $W$. The circuit repeats infinitely across space (abscissa) and time (ordinate), as well as all figures in the sequel. Notice that there are two light-like lattices (yellow and green) evolving independently.}
\label{fig:LLLat1}
\end{figure}

\noindent {\em Usual QWs} are over the space $\ell^2(\varepsilon\Z; \C^{k}\oplus \C^{k})$. We write $\phi(t)$ for a field taking a lattice position $x$ into $\C^{2k}$-vector. These QWs are obtained through the repeated application of a local unitary $W$ from $\C^{2k}$ to $\C^{2k}$, referred to as the `coin'. Hence $2k$ is the coin dimension or internal degree of freedom of the walker. The reason why it splits as $(k+k)$ is because each $W(t,x)$ takes the $k$ upper components of $\phi(t,x-2\varepsilon)$ and the $k$ lower components of $\phi(t,x+2\varepsilon)$, in order to produce $\phi(t+2\varepsilon,x)$.\\
Therefore the inputs and outputs of the different $W(t,x)$ are non-overlapping and the single-step evolution operator of the QW writes
\begin{equation*}
U(t) := \bigoplus_{x \in \varepsilon\Z} W(t,x)
\end{equation*}
where $t$ indicates the possible time dependence of the local unitaries. Therefore usual QWs evolve two independent light-like lattices, as made clear in Fig. \ref{fig:LLLat1}. On one of the light-like lattices, the evolution is given by 
\begin{small}
\begin{equation*}
V(t) := \bigoplus_{x \in 2\varepsilon\Z} W(t,x)\text{~and~}V(t+2\varepsilon) :=  \!\!\bigoplus_{x \in 2\varepsilon\Z + \varepsilon}\!\! W(t+2\varepsilon,x).
\end{equation*}
\end{small}
whilst on the other lattice everything is shifted in position.

\begin{figure}[h!]
\includegraphics[width=\columnwidth]{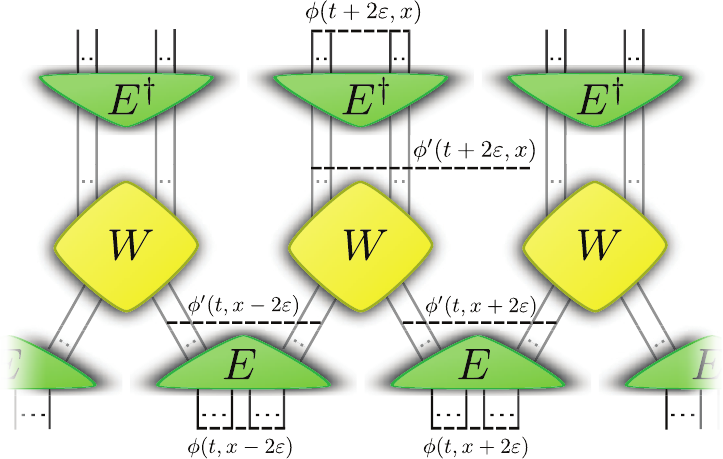}
    \caption{(Color Online) Each $\phi$ groups the $\psi$ fields at $2k$ locations (black solid wires). Then, $\phi$ is encoded via $E$ to obtain $\phi'$, which the operator $W$ acts upon. Finally, a decoding step is performed with $E^\dagger$.}
\label{fig:LLLat2}
\end{figure}

\noindent {\em Paired QWs} were introduced in \cite{ArrighiGRDirac,ArrighiGRDirac3D}. In the context of this paper they arise as follows. Consider a scalar field $\psi:\delta\mathbb{Z}\rightarrow\mathbb{C}$ such that $k\delta=\varepsilon$.  Let $\phi$ group the $\psi$ at $2k$ successive locations, ignoring one in two sites for convenience (see later Eq.\eqref{eq:alphadelta}):
\begin{small}
\begin{align}
&\phi(t,x)=\frac{1}{\sqrt{2k}}\big(\ldots,\psi(t,x-\delta),\psi(t,x+\delta),\ldots\big)\label{eq:grouping}\\
&\textrm{i.e. }\phi(t,x)_{j} =\frac{\psi(t,x+2j\delta)}{\sqrt{2k}}\quad\textrm{with}\quad j\in -k+1\ldots k\\
\end{align}
\end{small}
Then, apply a unitary encoding $E$ to each group, we obtain $\phi'(t,x)=E\phi(t,x)$. Finally, define a QW over the space $\bigoplus_{2\varepsilon\mathbb{Z}} (\C^{k}\oplus\C^{k})$ of these encoded groups $\phi'$. The local unitary $W$ will be from $\C^{2k}$ to $\C^{2k}$, and each $W(t,x)$ will take the $k$ upper components of $\phi'(t,x-2\varepsilon)$ and the $k$ lower components of $\phi(t,x+2\varepsilon)$ in order to produce $\phi'(t+2\varepsilon,x)$. The inputs and outputs of the different $W(t,x)$ are again non-overlapping and they can be applied synchronously to generate the QW evolution over the full space lattice, 
\begin{equation*}
U(t) := \bigoplus_{x \in 2\varepsilon\Z} W(t,x).
\end{equation*}
In the end, each $\phi'(t+2\varepsilon,x)$ is decoded as $\phi(t+2\varepsilon,x)=E^\dagger \phi'(t+2\varepsilon,x)$ and ungrouped as \eqref{eq:grouping}. Notice that this Paired QW (pictured in Fig. \ref{fig:LLLat2}), phrased in terms of $\phi'$, is therefore but a subcase of the usual QW definition---from a discrete point of view at least.\\ 
When taking the continuum limit, however, a subtle difference shows up. Indeed, the regularity of initial condition is given in terms of $\psi(t)$, which here assumed to be a smooth scalar function, i.e. $\psi(t,x)\approx\psi(t,x+\delta)$. It follows that the grouping $\phi(t)$ will be smooth both externally, i.e. $\phi(t,x)\approx\phi(t,x+\varepsilon)$, and internally, i.e. $\phi(t,x)\approx\bigoplus_{2k}\psi(t,x)$, which is not so usual to ask for. These reinforced regularity conditions are key to richer continuum limits.\\
Altogether, we have a quantum walk of the kind shown in Fig. \ref{fig:LLLat2}.

\noindent {\em Focussing on a gate} is all that is needed in order to work out the continuum limit. Again, the unitary coin $W$ receives $k$ wires from the left and $k$ wires from the right. Each of the left (resp. right) $k$ wires are obtained as the right (resp. left) projection $P^+$ (resp. $P^-$) of the $2k\times 2k$ unitary encoding $E$ applied to $\phi(t,x-2\varepsilon)$ (resp. $\phi(t,x+2\varepsilon)$). The output to $W$ then gets decoded with $E^\dagger$. Hence at $(x,t)$ the relevant input is made of $\phi(t,x-2\varepsilon)$ and $\phi(t,x+2\varepsilon)$ and the relevant output is $\phi(t+2\varepsilon,x)$.\\
Altogether, let $X=\sigma_x\otimes I_k$, and $C= C_0 e^{i \varepsilon H}$, where $C_0=WX$ and $H$ is a hermitian perturbation, we have:
\begin{align}
\phi(t+2\varepsilon,x)=&E(t+2\varepsilon)^\dagger C\\ 
&\big( P^+ E(t,x-2\varepsilon)\phi(t,x-2\varepsilon)\\ 
+&P^-E(t,x+2\varepsilon)\phi(t,x+2\varepsilon) \big)\label{eq:themodel}
\end{align}
(Because we looked at $\phi(t,x \pm 2\varepsilon)$ and wish the lattice to be lightlike, we consider that $2\varepsilon$ units of time have past.)

\section{Fixed speed case: continuum limit}\label{sec:fixedspeedconditions}

In order to look at the continuum limit of the quantum walk, we will assume that $\psi$ is differentiable. In fact, only its first order expansion will matter to this continuum limit. We have
\begin{align}
\phi(t,x)_{j} &=\frac{1}{\sqrt{2k}}\left[\psi(t,x)+\frac{(2j+1)}{k}\varepsilon \partial_x\psi(t,x)\right]\\
\phi&=\psi\ket{\alpha}+\varepsilon \partial_x\psi\ket{\delta}\\
\textrm{with}\quad\ket{\alpha} &= \frac{1}{\sqrt{2k}}(1, 1, 1,\dots), \\
\ket{\delta} &= \frac{1}{k\sqrt{2k}}(-2k+1, \ldots, -1, 1 \dots, 2k-1).
\end{align}
Let us stress again that $\psi$ is a complex scalar thus that $\psi\ket{\alpha}$ and $\partial_x\psi\ket{\delta}$ should be understood as scalar multiplication respectively with $\ket{\alpha}$ and $\ket{\delta}$, which each have dimension $2k$ and
\begin{align}
\braket{\alpha}{\delta} &= 0 \label{eq:alphadelta}\\
\Vert \ket{\alpha} \Vert^2 &= 1 \\
\Vert \ket{\delta} \Vert^2 &= \frac{1}{k^3}\sum_{i=0}^{k-1} (1 + 2 i)^2\\
&= \frac{1}{3k^2} (4k^2-1) \\
&= \frac{4}{3}\left[1 - \frac{1}{(2k)^2}\right]
\end{align}
We have, up to $\varepsilon^2$:
\begin{align}
\phi(t,x\pm2\varepsilon)&=\psi\ket{\alpha}+\varepsilon \partial_x\psi\ket{\delta}\pm 2\varepsilon\partial_x\psi\ket{\alpha}\\
\phi(t+2\varepsilon,x)&=\psi\ket{\alpha}+\varepsilon\partial_x\psi \ket{\delta}+2\varepsilon \partial_t\psi \ket{\alpha}
\end{align}
Let
\begin{align}
\ket{\alpha'}&=E\ket{\alpha}\\
\ket{\delta'}&=E\ket{\delta}\\
C  &= W X 
\label{eq:Wdev}
\end{align}
and notice that in this section we consider $C \equiv C_0$.

Now we can expand the encoded output, i.e. what comes out of $C$ prior to the decoding $E^\dagger$. This is, up to $\varepsilon^2$ terms:
\begin{align}
\phi'_{out}=&\psi(\ket{\alpha'}+\varepsilon \partial_x\psi \ket{ \delta'}+2\varepsilon \partial_t\psi \ket{\alpha'}
\end{align}
Similarly we can expand the encoded inputs, i.e. what comes out of the $E$s and gets fed into $C$. \\
\
\begin{align}
&\phi'_{in}=\\
&P^+\big(\psi\ket{\alpha'}
+\varepsilon \partial_x\psi \ket{ \delta'}-2\varepsilon \partial_x\psi \ket{\alpha'}\big)\\
&P^-\big(\psi\ket{\alpha'}
+\varepsilon \partial_x\psi \ket{ \delta'}+2\varepsilon \partial_x\psi \ket{\alpha'}\big)
\end{align}

\subsection{$0^{th}$ order}

When $\varepsilon$ is zero, we need the (encoded) output to be equal to the (encoded) inputs, otherwise the field varies discontinuously in time and has no continuum limit. This is the zeroth order condition.\\
But in this case the (encoded) left-incoming and right-incoming parts of the input are equal. We can then use $P^+ v + P^- v = v$, so that the zeroth order condition becomes
\begin{align}
\ket{\alpha'} &= C\ket{\alpha'}. \label{eq:cond1}
\end{align}

\subsection{$1^{st}$ order.}
If Eq. \eqref{eq:cond1} is satisfied, then in Eq. \eqref{eq:themodel} all we are left with are the first order terms:
\begin{align}\label{eq:cond2}
&\partial_x\psi \ket{ \delta'}+2\partial_t\psi \ket{\alpha'}\\
&=C\\
&\big(P^+\big(\partial_x\psi \ket{ \delta'}-2\partial_x\psi \ket{\alpha'}\big)\\
+&P^-\big(\partial_x\psi \ket{ \delta'}+2\partial_x\psi \ket{\alpha'}\big)\big)\\
&=\partial_x\psi C\ket{ \delta'}+2 \partial_x\psi C Z\ket{\alpha'}
\end{align}
where we introduced $Z=\sigma_z\otimes I_k$ and used $-P^+ v + P^- v = Zv$.

Now, project it with $\ket{\alpha'}$:
\begin{small}
\begin{align}\label{eq:cond3}
\partial_x\psi \braket{\alpha'}{ \delta'}+2\partial_t\psi \braket{\alpha'}{\alpha'}
&= \partial_x\psi \bra{\alpha'}C \ket{ \delta'}+\\
2 \partial_x\psi \bra{\alpha'}CZ\ket{\alpha'}\\
\partial_t\psi &=\partial_x\psi \bra{\alpha'}Z\ket{\alpha'}\\
\textrm{i.e. }\partial_t\psi &=c\partial_x\psi\label{eq:Dirac}\\
\textrm{with }c&:=\bra{\alpha'}Z\ket{\alpha'}\label{eq:P1}
\end{align}
\end{small}
this simple transport equation describes what happens to one component of the WE in curved spacetime with constant metric. Notice that although several speeds can be implemented by a single $W$, these speeds do not form a continuum---a single $W$ can only implement a finite number of speeds.

Let us now verify the constraint equation introducing the projector $Q$ orthogonal to $\ket{\alpha'}$, such that $Q + \ket{\alpha'}\bra{\alpha'} = I$. Projecting \eqref{eq:cond2} with $Q$ and rearranging we have:
\begin{equation}
\partial_x \psi (C\ket{ \delta'} - \ket{ \delta'} + 2QCZ\ket{\alpha'})=0
\end{equation}
In general, we want
\begin{align}
C^\dagger\ket{ \delta'} &= \ket{ \delta'} + 2Q Z\ket{\alpha'} + O(\varepsilon) \label{eq:constraint1} 
\end{align}

In order to satisfy equation \eqref{eq:constraint1}, we just need that the right member has the same norm as $\ket{ \delta'}$: in this case we can always define the action of $C$ in such a way that the equation holds. Therefore:
\begin{small}
\begin{equation}
\braket{ \delta'}{ \delta'} = \braket{ \delta'}{ \delta'} + 2 (\bra{ \delta'}QZ\ket{\alpha'} + \bra{\alpha'}ZQ\ket{ \delta'}) + 4 \bra{\alpha'}ZQZ\ket{\alpha'} 
\end{equation}
\end{small}
giving
\begin{equation}
\Re(\bra{ \delta'}Z\ket{\alpha'}) = -\bra{\alpha'}ZQZ\ket{\alpha'}.
\end{equation}

\section{Fixed speed case : discretized metric}\label{sec:discretemetric}

\begin{figure}[t]
\centering
\includegraphics[width=0.8\columnwidth]{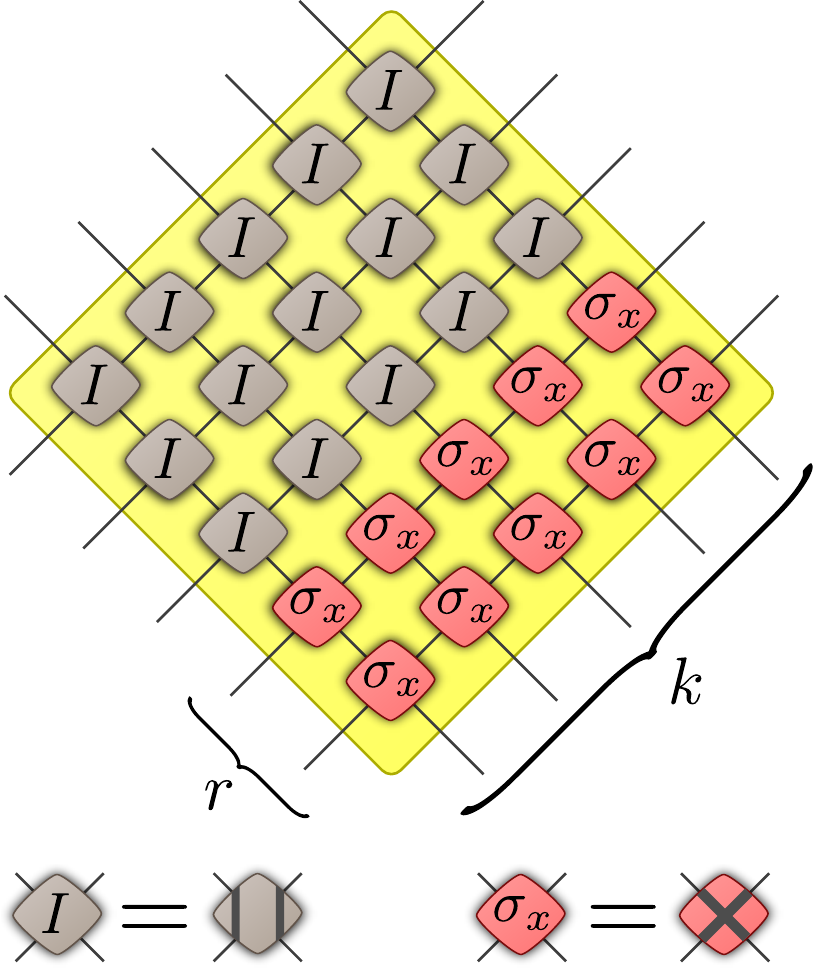} 
\caption{(Color Online) The internal structure of the $W$ operator in term of elementary gates $W_0 = I$ and $W_1=\sigma_x$, in the case $k=5$ and $r=2$. The identity gate corresponds to non-propagation, while the $\sigma_x$ corresponds to full-speed propagation.}
\label{fig:Wtiled}
\end{figure}

\begin{figure}[h!]
\includegraphics[width=\columnwidth]{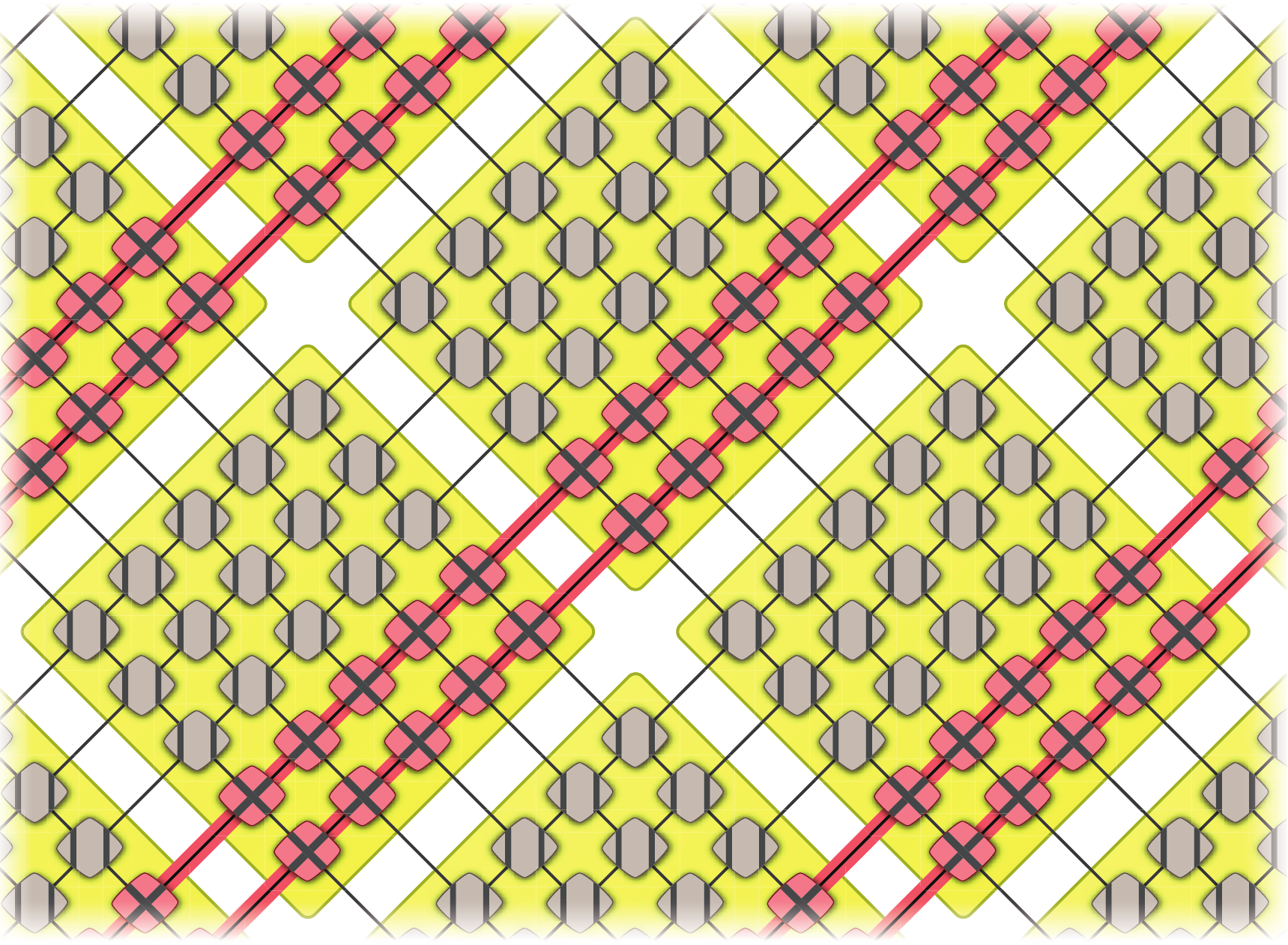} 
\caption{(Color Online) The background red signals, uniformly distributed with density $r/k$, activate the $\sigma_x$ gates as they traverse the lattice.}
\label{fig:WtiledSignals}
\end{figure}

We now proceed to study a discretized metric. We are interested in a model where the $W$ operator is composed of two kinds of elementary gates only, $W_0$ and $W_1$, as in Fig. \ref{fig:Wtiled}. With the goal of obtaining any possible velocity, a reasonable assumption is to take as elementary gates the extreme cases. Therefore we choose $W_0=I$, corresponding to non-propagation, and $W_1=\sigma_x$, corresponding to full-speed propagation.

Moreover we will assume that $W_1$ is activated only in response to special background signals (pictured in red in Fig. \ref{fig:WtiledSignals}), traversing the lattice at lightspeed. These signals can be seen as charge-less and massless scalar particles, uniformly distributed on the spacetime grid at initial time and we indicate with $r \leq k$ the number of signals per-tile (their density is therefore $r/k$).

The corresponding $C$ is of the form $C = C_B \oplus I_r$ where $C_B$ is of dimension $2k-r$ and acts as a right-shift of $k-r$ positions. The action on a generic vector $(x_0, \dots, x_{2k-1})^T$ is then the following:
\begin{equation}
C \begin{pmatrix}
  x_0 \\ \vdots \\ x_{2k-r-1} \\ \cline{1-1} x_{2k-r} \\ \vdots \\ x_{2k-1}
\end{pmatrix}
=
\begin{pmatrix}
  x_k \\ \vdots \\ x_{2k-r-1} \\ x_0 \\ \vdots \\ x_{k-1} \\ \cline{1-1} x_{2k-r} \\ \vdots \\ x_{2k-1}
\end{pmatrix}.
\end{equation}
We need to satisfy eq. \eqref{eq:cond1}, \eqref{eq:cond2} and the normalization conditions.
If we project equation  \eqref{eq:cond2} we have
\begin{align}
(C^\dagger_B - I_B)\ket{\delta'}_B &= 2(Z_B-cI_B)\ket{\alpha'}_B\label{eq:cond2B}\\
0 &= 2(Z_R-cI_R)\ket{\alpha'}_R  \label{eq:cond2R}
\end{align}
Since $Z_R=-I_R$, equation \eqref{eq:cond2R} gives immediately
\begin{equation}
(c+1) \ket{\alpha'}_R = 0
\end{equation}
so if we want non-trivial solutions $c\neq -1$ we have to set $\ket{\alpha'}_R = 0$.

We will look for solutions of the form
\begin{equation}\label{eq:al1}
\ket{\alpha'}_B = \begin{pmatrix}
  a \\ \vdots \\ a
\end{pmatrix},
\ket{\delta'}_B = \begin{pmatrix}
  d \\ d + \beta \\ d + 2\beta \\ \vdots \\ d + (2k-r-1)\beta
\end{pmatrix}
\end{equation}
for suitable real numbers $a, d, \beta$. This linear form for $\ket{\delta'}$ is suggested by the linearity of $\ket{\delta}$ and the form of $\ket{\alpha'}_B$ is the simplest choice of invariant vector under permutation of the components. \\

From the normalization of $\ket{\alpha'}$ we derive
\begin{equation}
a = \frac{1}{\sqrt{2k-r}}.
\end{equation}

Equation \eqref{eq:cond1} is then automatically satisfied.
Equation \eqref{eq:cond2B} becomes
\begin{equation}\label{eq:cond2Bsystem}
\begin{pmatrix}
d_{k-r} \\ \vdots \\ d_{2k-r-1} \\ \cline{1-1} d_0 \\ \vdots \\ d_{k-r-1}
\end{pmatrix}
-
\begin{pmatrix}
d_0 \\ \vdots \\ d_{k-1} \\ \cline{1-1} d_k \\ \vdots \\ d_{2k-r-1}
\end{pmatrix}
=
2a\begin{pmatrix}
1-c \\ \vdots \\ 1-c \\ \cline{1-1} -1-c \\ \vdots \\ -1-c
\end{pmatrix}
\end{equation}
where now the split is $k \oplus (k-r)$. This is equivalent to two independent equations:
\begin{align}
\beta(k-r)&=2a(1-c) \\
-\beta k&=2a(-1-c).
\end{align}
Solving,
\begin{align}
c &= \frac{r}{2k-r}\\
\beta &= \frac{4a}{2k-r} =  \frac{4}{(2k-r)^{3/2}}.\label{eq:defra}
\end{align}

Any possible value for $r\leq k<\infty$ thus induces a certain speed $c(k,r)$, and these are dense (in the same way that rational numbers are dense within real numbers) in $[-1,1]$. All of these speeds can be recovered by combining $W_0$ and $W_1$. Whilst it is well-known that in standard quantum walks, the speed of propagation is a cosine function of coins rotation angle, here the same range of speeds is recovered by a combination of two fixed coins.

From \eqref{eq:defra} it follows that:
\begin{align}\label{eq:defr}
r &= \frac{2kc}{1+c}\\\label{eq:defa}
a &= \sqrt{\frac{1+c}{2k}}.
\end{align}
Now, we just need to verify the normalization constraint for $\ket{\delta'}$. We need to determine $d$ such that
\begin{equation}\label{eq:condSubNorm}
\Vert \ket{\delta'}_B\Vert^2  \leq  \Vert \ket{\delta}\Vert^2,
\end{equation}
with
\begin{equation}
\Vert \ket{\delta'}_B\Vert^2 = \sum_{i=0}^{2k-r-1} (d+\beta i)^2.
\end{equation}
The minimum is obtained by differentiating the above and setting it to zero:
\begin{align}
0&= (2k-r)d+\beta\frac{(2k-r-1)(2k-r)}{2}\\
d&= -\frac{\beta}{2}(2k-r-1).
\end{align}
The minimum is then
\begin{small}
\begin{align}
\beta^2 \frac{1}{12}(2k-r-1)(2k-r)(2k-r+1) = \\
\frac{4}{3}\left[1 - \frac{1}{(2k-r)^2}\right].
\end{align}
\end{small}
It follows that
\begin{equation}
\frac{4}{3}\left[1-\frac{1}{(2k-r)^2}\right] < \frac{4}{3}\left[1-\frac{1}{(2k)^2}\right],
\end{equation}
proving the condition \eqref{eq:condSubNorm}, and thus the normalisation of $\ket{\delta'}$.

\section{Non-fixed speed case: continuum limit }\label{sec:nonfixedspeedcaseconditions}

We demand that $E$ be continuous, but not necessarily differentiable. It follows that $\ket{\alpha'}$ is continuous but not necessarity differentiable. We can define
\begin{align}
\ket{\alpha'(t,x\pm2\varepsilon)}&=\ket{\alpha'}\pm 2\varepsilon\ket{\Delta^{\pm}_x\alpha'}\\
\ket{\mu_x\alpha'}&=(\ket{\Delta^+_x\alpha'}-\ket{\Delta^-_x\alpha'})/2\\
\ket{\Delta_x\alpha'}&=(\ket{\Delta^+_x\alpha'}+\ket{\Delta^-_x\alpha'})/2\\
\ket{\alpha'(t,x\pm2\varepsilon)}&=\ket{\alpha'}\pm 2\varepsilon\ket{\Delta_x\alpha'}+2\varepsilon\ket{\mu_x\alpha'} \label{eq:alphaexp}
\end{align}
and similarly w.r.t time and over $\ket{\alpha'}$, $\ket{\delta'}$. Notice that $\ket{\mu_x\alpha'}$ measures the extent in which $E$ is non-differentiable when acting over $\ket{\alpha}$.
\begin{align}
E(t+2\varepsilon,x)\phi_{out}=&\psi(\ket{\alpha'}+2\varepsilon\ket{\Delta^+_t\alpha'})\\
&+\varepsilon \partial_x\psi \ket{\delta'}+2\varepsilon \partial_t\psi \ket{\alpha'}
\end{align}
\begin{align}
&(E(t,x-2\varepsilon)\oplus E(t,x+2\varepsilon))\phi_{in}=\\
&\psi(\ket{\alpha'}\oplus\ket{\alpha'})+\varepsilon \partial_x\psi (\ket{\delta'}\oplus\ket{\delta'})\\
&+2\varepsilon \partial_x\psi ((-\ket{\alpha'})\oplus\ket{\alpha'})
+2\varepsilon\psi((-\ket{\Delta_x\alpha'})\oplus\ket{\Delta_x\alpha'})\\
&+2\varepsilon\psi (\ket{\mu_x\alpha'}\oplus\ket{\mu_x\alpha'})
\end{align}

More in general in this section we will consider a non-vanishing perturbation $H$, then 
\begin{equation}
C = C_0 e^{i \varepsilon H} = C_0 + i \varepsilon C_0 H,
\label{eq:pert}
\end{equation}
hence we can define $C_1 = i C_0 H$.

\section*{$0^{th}$ order}

Again,
\begin{align}
\ket{\alpha'} &= C_0 \ket{\alpha'}\label{eq:cond1nfs}
\end{align}

\section*{$1^{st}$ order.}
If \eqref{eq:cond1} is satisfied, we are left with $1^{st}$ order terms:
\begin{equation}\label{eq:cond2NFS}
\begin{split}
&2\psi\ket{\Delta^+_t\alpha'}+\partial_x\psi \ket{\delta'}+2\partial_t\psi \ket{\alpha'}\\
&=W(P\oplus P')\big( \partial_x\psi (\ket{\delta'}\oplus\ket{\delta'})\\
&+2 \partial_x\psi ((-\ket{\alpha'})\oplus\ket{\alpha'})\\
&+2\psi((-\ket{\Delta_x\alpha'})\oplus\ket{\Delta_x\alpha'})\big) +2\psi (\ket{\mu_x\alpha'}\oplus\ket{\mu_x\alpha'}) \big)\\
&=\partial_x\psi C_0\ket{\delta'}+2 \partial_x\psi C_0 Z\ket{\alpha'}+2\psi C_0 Z\ket{\Delta_x\alpha'}\\
&+ 2\psi C_0 \ket{\mu_x \alpha'} + \psi  C_1 \ket{\alpha'}
\end{split}
\end{equation}
Now, project it with $\ket{\alpha'}$:
\begin{align}\label{eq:cond2Palpha}
\partial_t\psi =& c\partial_x\psi
+\psi \bra{\alpha'}Z\ket{\Delta_x\alpha'} \\
&+\psi (\braket{\alpha'}{\mu_x \alpha'}-\psi\braket{\alpha'}{\Delta^+_t\alpha'} + \bra{\alpha'}C_1\ket{\alpha'}
\end{align}
and in order to recover the good equation we require $\bra{\alpha'}C_1\ket{\alpha'}$ = 0.\\ 
Now, let us define
\begin{align}
m&=\bra{\alpha'}Z\ket{\Delta_x\alpha'}\\
n&=-\braket{\alpha'}{\Delta^+_t\alpha'}\\
s&=\braket{\alpha'}{\mu_x\alpha'}.
\end{align}
We call `differentiable' the case where $\ket{\mu_x\alpha'}=0$. In this limit we have that \eqref{eq:cond2Palpha} becomes the differential equation
\begin{equation}\label{eq:ContLimitEq}
\partial_t\psi
=c\partial_x\psi+ (m + n)\psi.
\end{equation}
In order to show that this is the WE in curved spacetime, we need to prove that
\begin{equation}\label{eq:rem}
\Re(m) =\frac{1}{2}\partial_x c
\end{equation}
Indeed,
\begin{align}
\frac{c(x+2\varepsilon) - c}{2\varepsilon} = \bra{\Delta_x\alpha'} Z\ket{\alpha'} + c.c. + O(\varepsilon)
\end{align}
When $\varepsilon \to 0$ this becomes
\begin{equation}
\partial_x c = m+m^*
\end{equation}
proving \eqref{eq:rem}.\\
Moreover, since
\begin{align}
0 &= \partial_t \braket{\alpha'}{\alpha'} = \lim_{\varepsilon\to 0} \frac{\braket{\alpha'(t+2\varepsilon)}{\alpha'(t+2\varepsilon)} - \braket{\alpha'}{\alpha'}}{2\varepsilon} \\
&=  \lim_{\varepsilon\to 0} \frac{(\bra{\alpha'} + 2\varepsilon \bra{\Delta^+_t\alpha'})(\ket{\alpha'} + 2\varepsilon \ket{\Delta^+_t\alpha'}) - \braket{\alpha'}{\alpha'}}{2\varepsilon}\\
&= -(n+n^*)
\end{align}
it follows that $n$ is imaginary. Therefore we can write \eqref{eq:ContLimitEq} as
\begin{equation}
\partial_t\psi
=c\partial_x\psi+\frac{1}{2}(\partial_x c)\psi+(\ii\Im(m)+n)\psi,
\end{equation}
which is a non-conservative first-order quasilinear hyperbolic equation. 

When $\ket{\alpha'}$ is real this is
\begin{align}
\partial_t\psi &=c\partial_x\psi+\frac{1}{2}(\partial_x c)\psi
\end{align}
which coincided with the (1+1)-dimensional WE in curved spacetime, looking at only one component of the spinor.

In general, the norm is conserved if and only if
\begin{align}
\lim_{\varepsilon\rightarrow 0} \Re(m+n+s)=\frac{1}{2}\partial_x c
\label{eq:condNorm}
\end{align}
In the next section we will show that the tiled model is non-differentiable ($s\neq 0$) and the previous equation is not satisfied.

Moreover, in order to obtain the generalised constraint equations,  we need to project \eqref{eq:cond2NFS} with $Q$ and, by rearranging all the terms,  we have:
\begin{small}
\begin{align}
\psi (Q \ket{\Delta_t^+\alpha'} - QC_0Z \ket{\Delta_x\alpha'} - QC_0 \ket{\mu_x \alpha'}) + \psi Q C_1 \ket{\alpha'} + \\
\partial_x \psi (C_0\ket{ \delta'} - \ket{ \delta'} + 2QC_0Z\ket{\alpha'}) =0.
\end{align}
\end{small}
And, since $\psi$ and $\partial_x\psi$ have to be independent, we consider separately the following two equations:
\begin{small}
\begin{align}
\Gamma_1 &= Q \ket{\Delta_t^+\alpha'} - QC_0Z \ket{\Delta_x\alpha'} -
 QC_0 \ket{\mu_x \alpha'} +  Q C_1 \ket{\alpha'} =0 \label{CNFS1}\\
 \Gamma_2 &= C_0\ket{ \delta'} - \ket{ \delta'} + 2QC_0Z\ket{\alpha'} =0.
\end{align}
\end{small}

\section{Non-fixed speed case \& discretized metric: extra terms}\label{sec:nonfixedspeeddiscretizedmetric}

Now, let us see if the tiled model works for non-constant speed $c$.
We have from equations \eqref{eq:defr} and \eqref{eq:defa} that
\begin{align}\label{eq:derar}
\partial _x r &= \frac{2k\partial_x c}{(1+c)^2}\\
\partial _t r &= \frac{2k\partial_t c}{(1+c)^2}\\
\partial_c a &= \frac{1}{2\sqrt{2k(1+c)}},
\end{align}
where we are assuming that $c(t,x)$ is regular, continuous and at least once differentiable. 
Suppose that $E$ depends now on $c(t,x)$. 
From equation \eqref{eq:al1}:
\begin{align}
&\ket{\alpha'(c(t,x))}=\bigoplus_{2k-r}a(c(t,x))\bigoplus_{r}0\\
&\ket{\alpha'(c(t,x+2\varepsilon))}=\bigoplus_{2k-r-2\varepsilon\partial_x r}a(c(t,x+2\varepsilon))\bigoplus_{r+2\varepsilon\partial_x r} 0\\
&\ket{\alpha'(c(t,x-2\varepsilon))}=\bigoplus_{2k-r+2\varepsilon\partial_x r}a(c(t,x-2\varepsilon))\bigoplus_{r-2\varepsilon\partial_x r} 0\\
&\ket{\alpha'(c(t+2\varepsilon,x))}=\bigoplus_{2k-r-2\varepsilon\partial_t r}a(c(t+2\varepsilon,x))\bigoplus_{r+2\varepsilon\partial_t r} 0
\end{align}

And, we derive the finite differences:

\begin{align}
\ket{\Delta_t^+\alpha'} &= \frac{\ket{\alpha'(c(t+2\varepsilon,x)} - \ket{\alpha'(c(t,x)} }{2\varepsilon} \\
\ket{\Delta_x\alpha'} &= \frac{\ket{\alpha'(c(t,x+2\varepsilon)} - \ket{\alpha'(c(t,x-2\varepsilon)} }{4\varepsilon} \\
\ket{\mu_x\alpha'} &= \frac{\ket{\alpha'(c(t,x+2\varepsilon)} + \ket{\alpha'(c(t,x-2\varepsilon)} - 2 \ket{\alpha'} }{4\varepsilon}
\end{align}

and in particular for the tiled model and assuming, $\partial_r c = 0$, \textit{i.e.} $\partial_x r =  -\partial_t r$, we get:
\begin{small}
\begin{align}
&\ket{\Delta^+_t\alpha'} = \bigoplus_{2k-r} (-\partial_c a \partial_x c) \bigoplus_{2\varepsilon \partial_x r} (\frac{a}{2\varepsilon} - \partial_c a \partial_x c)\bigoplus_{r - 2\varepsilon \partial_x r} 0\\
&\ket{\Delta_x\alpha'} = \bigoplus_{2k-r-2\varepsilon\partial_x r}   \partial_c a \partial_x c \bigoplus_{4\varepsilon\partial_x r} (\frac{a}{4\varepsilon} -\frac{ \partial_c a \partial_x c}{2\varepsilon})  \bigoplus_{r-2\varepsilon\partial_x r} 0\\
&\ket{\mu_x\alpha'} = \bigoplus_{2k-r-2\varepsilon\partial_t r} 0 \bigoplus_{2\varepsilon\partial_x r}  (-\frac{a}{4\varepsilon} -\frac{ \partial_c a \partial_x c}{2\varepsilon})   \bigoplus_{2\varepsilon\partial_x r} (\frac{a}{4\varepsilon} -\frac{ \partial_c a \partial_x c}{2\varepsilon})\\
&\bigoplus_{r- 2 \varepsilon \partial_x r} 0
\end{align}
\end{small}

Now we can compute each term of equation \eqref{eq:condNorm}:

\begin{align}
s &= - \frac{\partial_x c}{2(1+c)}\\ 
n &= \frac{\partial_x c}{2(1+c)}\\
m &= \frac{1}{2} \partial_x c,
\end{align}
and conclude that, because \eqref{eq:condNorm} is satisfied, Eq. \eqref{eq:cond2Palpha} coincides with:
\begin{equation}\label{eq:CDE}
\partial_t\psi =c\partial_x\psi+ \frac{\partial_x c}{2} \psi.
\end{equation}

Moreover we need to satisfy the constraints on $\Gamma_1$ and on $\Gamma_2$. A straightforward computations tells us that one on $\Gamma_2$ is automatically satisfied. The equation on $\Gamma_1$ reads:
\begin{small}
\begin{equation}\label{eq:gamma1}
\Gamma_1 = \bigoplus_{k-r} m a \bigoplus_{k} (- 4 \partial_c a \partial_x c + m a) \bigoplus_{r} 0  - Q C_1 \ket{\alpha'} =0.
\end{equation}
\end{small}

In order to satisfy the above equation, we don't want to choose any family of perturbation, but only those ones which could be decomposed locally on our circuit. One simple choice is to let the local perturbations act only on the border. Everywhere we apply $X$, except on the border, where the perturbation $\exp(i H(\theta) \varepsilon)$ is a rotation depending on a family of parameters $\theta^{i}_t$. 
\begin{figure}[t]
\centering
\includegraphics[width=\columnwidth]{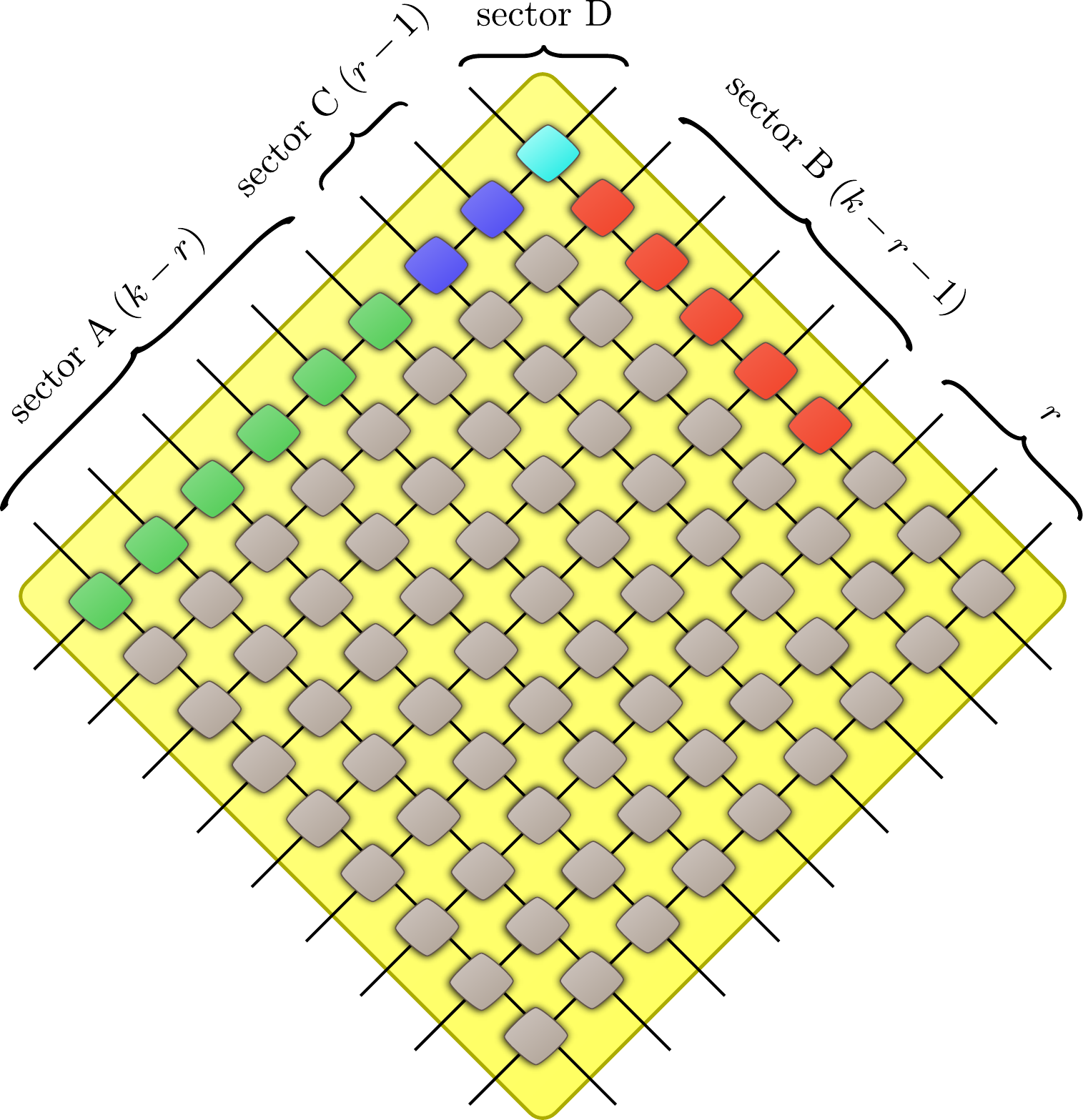} 
\caption{(Color Online) \label{fig:perturbations}Local perturbations, applied on the border of the $W$ operator. In sector (A), Green, $R(\theta^{(1)}_t)$; in sector (B), Orange, $R'(\theta^{(3)}_t)$; in sector (C), Blue, $R(\theta^{(2)}_t)$; in sector (D), Cyan, $R(\theta^{(4)}_t)$.}
\label{fig:pert}
\end{figure}

As we can see in Fig. \ref{fig:perturbations}, our choice is to apply in sector (A), the local perturbation $R(\theta^{(1)}_t)$, where $t = 1\dots k-r $; in the sector (C), $R'(\theta^{(2)}_t)$, where $t = 1\dots k-r-1$, in the sector (B), $R(\theta^{(3)}_t)$, where $t = 1\dots r-1$ and finally in the sector (D), at the junction between sector (B) and (C), we apply $R(\theta^{(4)}_t )$. The matrices and the angles are defined as follows:
\begin{align}
    R(\theta^{(i)}_t) &= \begin{pmatrix}\cos\theta^{(i)}_t & \sin\theta^{(i)} \\  -\sin\theta^{(i)}_t & \cos\theta^{(i)}_t \end{pmatrix} \\
    R'(\theta^{(i)}_t) &=  R(\theta^{(i)}_t+\pi/2) \\
\theta^{(1)}_t &= -\varepsilon m t\\
\theta^{(2)}_t &=  \varepsilon \frac{r-k}{k} m t \\
\theta^{(3)}_t &= -\varepsilon m \frac{r-k}{k}(1-k)t\\ 
\theta^{(4)}_t &= - \varepsilon m \frac{(r-k)^2}{k}
\end{align}
With this choice, the equation \eqref{eq:gamma1} is then satisfied. 

\section{Summary, Discussion and conclusion}\label{sec:conclusion}

{\em Summary of results.} Spacetime discrete quantum mechanics provides both quantum simulation algorithms of physical phenomena, and neat toy models for these. Quantum Walks (QW), in particular, are succesfully being used to simulate fundamental physics equations on the lattice---via the application of a local unitary coin $W$ across space. In this work we considered a very wide class of QW, referred to as Grouped QW, and thouroughly investigated their ability to simulate the equation
\begin{equation}
\partial_t\psi =c\partial_x\psi+\frac{1}{2}(\partial_x c)\psi,
\label{eq:scalarpropagation}
\end{equation}
i.e. the transport equation of the scalar $\psi$ in $(1+1)-$dimensional curved spacetime in abitrary coordinates, from which one can easily build the Weyl equation, which in turn is the basic ingredient of the Dirac equation.\\
We characterized the grouped QW that have Eq. \eqref{eq:scalarpropagation} as their continuum limit, by means of two equations \eqref{eq:cond1} and \eqref{eq:constraint1}. In the fixed speed case (i.e. when $c$ is not a function of $x$ and $t$), we gave a construction to generate every possible solution. In other words, given a certain speed $c$, the construction provides a way to generate every possible coin $W(c)$ that simulates the equation.\\
We them moved on to the question whether this was possible using only a finite number of coins---in order to simplify any potential experimental implementations of the QW, and as a mean to evaluate the ``ultimate discreteness of the metric hypothesis'' dear to several approaches of quantum gravity. In other words, could there be a single, fixed coin $W$ that is able to simulate \eqref{eq:scalarpropagation} for any given $c$? Perhaps by encoding the requested $c$ in the initial wave-packet in some way? We proved that a single $W$ can only serve to simulate a finite number of possible speeds, bounded by $\dim(W)-1$. But we also proved that there are two precise coins, $W_0$ and $W_1$, which can be combined as diamond-shaped circuits to form larger coins $W'(c)$, so that these larger coins can simulate any speed in real-time. Moreover, the prescription of the diamond can be encoded in the initial state via some lighlike propagating background signals, which control which of $W_0$ or $W_1$ should be applied. Therefore, a finite number of coins suffices to simulate any speed $c$.\\
In curved spacetime the speed of propagation may vary, however (i.e. $c$ is a function of $x$ and $t$). In the continuous setting, it is assumed to vary in a differentiable manner. In the discrete setting, if we allow ourselves a continuum of possible coins $W(c)$, then we can also engineer these so that the function $c\mapsto W(c)$ be differentiable. That way we can simulate the Eq. \eqref{eq:scalarpropagation} even in the non-fixed speed case. If, instead, we restrict ourselves to constructing diamond-shaped $W'(c)$ from a finite number of coins $W_i$, then it is likely that the function $c\mapsto W'(c)$ is non-differentiable. At least this was the case in every scheme we tried. This non-differentiabily introduces an extra term in the continuum limit, compared to \eqref{eq:scalarpropagation}. In some interesting cases this extra term vanishes---namely whenever $c$ is constant along a lightlike direction. We are then back to recovering \eqref{eq:scalarpropagation} as our continuum limit, but the first order of the contraint equation \eqref{eq:constraint1} fails. Fortunately, we showed that this could be fixed by means of simple, local rotations, thereby providing a convenient quantum simulation scheme. In our scheme, however, the number of local rotations required is no longer bounded as where the $W_i$, and they can no longer be controlled by lighlike propagating background signals. We leave it open whether a scheme exists that maintains these two nice properties.

\noindent {\em Interpretation of the results.} On the one hand, this paper provides concrete techniques to tune the propagation of a Quantum Walker, whilst making use of the least number of coin operators, differently from all previous models, in which the quantum coins were depending continuously upon the spacetime metrics. We hope that these will contribute to the QW toolbox of primitives that are used to formulate quantum algorithms and quantum simulation algorithms.\\
On the other hand, this paper has tackled the question whether discrete spacetime quantum mechanics is able to simulate propagation in curved spacetime, in the continuum limit. This question is central to Quantum Gravity, where a fundamentally discrete spacetime structure is often assumed or argued for and then quantized, but always with the hope to recover propagation of matter in countinuous curved spacetime in limit. This paper evaluated this question just for lattice-with-defects kinds of spacetime in $(1+1)-$dimensions, and aiming just at the curved spacetime scalar transport equation. Yet, this analysis has already revealed a deep concern: the non-differentiability of the discrete spacetime structure leads not to a traditional transport equation, but to a continuum limit equation that has an extra term accounting for this non-differentiability.\\
{\em Perspectives.} The Dirac equation in curved spacetime comes generalizing flat-space the Dirac equation to non-flat space, via a diffeomorphism. Perhaps it is therefore limited to a differentiable metric field, almost by definition/construction. We may wonder whether the equation could be generalized to non-differentiable curved spacetime instead---and perhaps this generalized equation is the only we may hope to recover using discrete spacetime structure.\\
Another approach would be to consider a wider class of discrete spacetime structures than those of this paper (graphs, spin networks). The quantum superpositions of these discrete spacetime structures could also be a key ingredient, which reintroduces diffentiability via the continuum of the amplitudes of the superpositions. 

\section*{Acknowledgements} This work has been funded by the ANR-12-BS02-007-01 TARMAC grant, the STICAmSud project 16STIC05 FoQCoSS and partially supported by the Spanish Ministerio de Educaci\'{o}n e Innovaci\'{o}n, MICIN-FEDER project FPA2014-54459-P, SEV-2014-0398 and Generalitat Valenciana grant GVPROMETEOII2014-087.

\appendix

\section{Fixed speed case : solutions}\label{sec:fixedspeedcasesolutions}

\subsection{Necessary conditions}

We use \eqref{eq:Dirac} in order to simplify Equation (\ref{eq:cond2}) as follows: 
\begin{equation}\label{eq:condfixedspeed2}
(C^\dagger-I)\ket{\delta'}=2(Z-cI)\ket{\alpha'}.
\end{equation}

Consider now the vector $\ket{o'}$, defined as 
the projection of $Z\ket{\alpha'}$ on the subspace orthogonal to Span$\{\ket{\alpha'}, \ket{\delta'}\}$. We can  express $Z\ket{\alpha'}$ on the $\{\ket{\alpha'}, \ket{\delta'}, \ket{o'}\}$ basis as follows:
\begin{equation}\label{eq:Zab}
Z\ket{\alpha'} = c\ket{\alpha'} + f \ket{\delta'} + \ket{o'},
\end{equation}
for suitable $f$.
Now, from this expression and \eqref{eq:constraint1} we also have
\begin{equation}\label{eq:Wab}
C^\dagger \ket{\delta'} = (2f+1)\ket{\delta'} + 2 \ket{o'}
\end{equation}
It will be useful to introduce the vector $\ket{\delta''}$ such that
\begin{align}
c\ket{\alpha'}+\ket{\delta''}&=Z\ket{\alpha'},\\
\textrm{i.e. }\ket{\delta''}&=(Z-cI)\ket{\alpha'}.\label{eq:deltasecond}
\end{align}
This is orthogonal to $\ket{\alpha'}$ by eq. \eqref{eq:P1}.
Now, taking norms in \eqref{eq:Zab} we have
\begin{align}
|f|^2\Vert \ket{\delta'}\Vert^2 + \Vert \ket{o'}\Vert^2 = \Vert \ket{\delta''}\Vert^2 \nonumber \\
=\Vert (Z-cI)\ket{\alpha'}\Vert^2 = 1-c^2 \label{eq:norm1}
\end{align}
implying the necessary condition
\begin{equation}\label{eq:norm1bis}
|f|^2 \leq \frac{\Vert \ket{\delta''}\Vert^2}{\Vert\ket{\delta'} \Vert^2} = 3 k^2 \frac{1-c^2}{4k^2-1}.
\end{equation}
Taking norms in \eqref{eq:Wab}  we have
\begin{align}
\vert 2f +1 \vert^2 \Vert\ket{\delta'}\Vert^2 + 4 \Vert\ket{o'}\Vert^2 =\Vert\ket{\delta'}\Vert^2 \label{eq:norm2}
\end{align}
Now, expanding and using \eqref{eq:norm1} this becomes
\begin{equation}\label{eq:norm2bis}
\Re(f)= -\frac{\Vert \ket{\delta''}\Vert^2}{\Vert\ket{\delta'}\Vert^2} = -3k^2\frac{1-c^2}{4k^2-1}.
\end{equation}
Then, the condition \eqref{eq:norm1bis} can also be expressed as
\begin{equation}\label{eq:norm1im}
\Im(f)^2 \leq \widetilde{f}^2
\end{equation}
with
\begin{equation}
\widetilde{f}^2 = \frac{\Vert\ket{\delta''}\Vert^2}{\Vert\ket{\delta'}\Vert^2} \left(1- \frac{\Vert\ket{\delta''}\Vert^2}{\Vert\ket{\delta'}\Vert^2} \right).
\end{equation}

\subsection{Building the solutions}

To find a valid $\ket{\alpha'}$, take any $r$, $l$, $\ket{0}$, $\ket{1}$ such that $Z\ket{0}=\ket{0}$, $Z\ket{1}=-\ket{1}$, $c=|r|^2-|l|^2$, $|r|^2+|l|^2=1$ and define:
\begin{align}
\ket{\alpha'}=r\ket{0}+l\ket{1} 
\end{align}
Then $\bra{\alpha'}Z\ket{\alpha'}=r^2-l^2= c$ as required by \eqref{eq:P1}. Notice that there is a $U(k)\times U(k)$ freedom in the choice of $\ket{\alpha'}$.

The vector $\ket{\delta''}$ is now also determined by $\ket{\alpha'}$ via eq. \eqref{eq:deltasecond}.

Choose $f$ compatible with equations \eqref{eq:norm2bis} and \eqref{eq:norm1im}.  We need a $\ket{\delta'}$ of norm $\Vert \ket{\delta}\Vert$ satisfying \eqref{eq:Zab}. Equivalently, we need that 
\begin{equation}\label{eq:prod}
\braket{\delta'}{\delta''} = f \Vert\delta'\Vert^2
\end{equation}
 from Eqs. \eqref{eq:Zab} and \eqref{eq:deltasecond}. But this is possible precisely because $f$ was chosen such that $|\braket{\delta'}{\delta''}| \leq \Vert\delta'\Vert \cdot \Vert\delta''\Vert$. 
Finally, take
\begin{equation}
\ket{\delta'''} = 2\ket{\delta''} + \ket{\delta'}
\end{equation}
By eq. \eqref{eq:norm2bis} and \eqref{eq:prod} it has norm $\Vert\ket{\delta'}\Vert$.

Now we pick any $C$ such that equation \eqref{eq:cond1} is satisfied, and such that
\begin{align}
C \ket{\delta'''} = \ket{\delta'},
\end{align}
the two combined leading to \eqref{eq:cond2}.

\

\subsection{Can several speeds coexist in a $C$?}

\subsection*{A number of speeds can coexist\ldots}

Can $c=-1$ and $c=1$ coexist in a $C$? Let us define $\ket{\alpha'_-}$ and $\ket{\alpha'_+}$ as the particular state $\ket{\alpha'}$ for $c=-1$ and $c=1$, respectively. We need that $\ket{\alpha'_+}$ and $\ket{\alpha_0'}_{-1}$ respect the following condition:
\begin{align}\label{eq:condAL}
C\ket{\alpha'_+} = \ket{\alpha'_+} \quad \text{and}\quad
C\ket{\alpha'_-} = \ket{\alpha'_-}.
\end{align}
It follows from equation \eqref{eq:P1} that $Z\ket{\alpha'_+} = \ket{\alpha'_+}$ and $Z\ket{\alpha'_-} = -\ket{\alpha'_-}$. How to choose $\ket{\delta'_\pm}$? From \eqref{eq:cond2} we have immediately that $C \ket{\delta'_\pm}$ = $\ket{\delta'_\pm}$. Thus, notice that the most unrestrictive choice for $C$ is $\ket{\delta'_+} = \ket{\alpha'_-}$ and $\ket{\delta'_-} = \ket{\alpha'_+}$, leaving $C = I_2 \oplus U_{2k-2}$. This choice satisfies \eqref{eq:cond1} and \eqref{eq:cond2} for both speeds.

Can arbitrary speeds coexist in a $C$? Consider several solutions $ c_i, E_i, C_i$ each of dimension $2k_i$, obtained applying the previous construction, but using everywhere $k'=\sum_i k_i$ instead of $k_i$. Arrange the $C_i$'s in blocks: $C=\bigoplus_i C_i$. Then, the same $C$ can be used for different speeds, with encodings $F_i$ defined as $E_i$ plus an embedding placing the encoded vector in the $i^{th}$ subspace. 

\subsection*{\ldots this number has to be finite}

We want to pick $\ket{\alpha'}$, and $\ket{\delta'}$ so that eq. \eqref{eq:cond2} is satisfied.
Let $P$ be the projector on the identity subspace of $C$ and $P^\perp$ the orthogonal projector. Let's write $\ket{\delta'} = P\ket{\delta'} + P^\perp \ket{\delta'}$ in eq. \eqref{eq:cond2}, which gives 
\begin{align}
(C^\dagger - I)(P\ket{\delta'} + P^\perp \ket{\delta'}) &= 2 (Z - cI)\ket{\alpha'} \\
(C^\dagger - I)P^\perp \ket{\delta'} &= 2 (Z - cI)\ket{\alpha'} \\
\end{align}
Projecting this equation with $P$ we have
\begin{align}
0 &= 2P(Z-cI)\ket{\alpha'} \\
PZ\ket{\alpha'} &= c\ket{\alpha'}
\end{align}
meaning that a necessary condition to have speed $c$ is that $c$ is an eigenvalue of the operator $PZ$. But the number of eigenvalues is finite.

\bibliography{biblio}
\bibliographystyle{plainurl}
\end{document}